\documentclass[twocolumn,a4,showpacs,preprintnumbers,amsmath,amssymb]{revtex4-1}
\usepackage{longtable}
\usepackage{epsfig}
\usepackage[colorlinks=true]{hyperref}

 \def\ket{\!>\,} \def\ack{\,|\,}

\usepackage{graphicx}
\usepackage{dcolumn}
\usepackage{bm}
\usepackage{mathptmx, courier, pifont}
\usepackage[scaled=1.0]{helvet}
\usepackage[T1]{fontenc}
\usepackage{textcomp}
\usepackage{color}
\usepackage{colordvi}
\usepackage[T1]{fontenc}

\begin{document}

\title{Quasiparticle and $\gamma$-band structures in $^{156}$Dy  }

\author{ S.~Jehangir$^{1,2}$, G.H.~Bhat$^{1,3,4}$, J.A.~Sheikh$^{1,4}$, S.~Frauendorf$^{5}$, S.N.T.~Majola$^{6,7}$, P.A.~Ganai$^{2}$ and J.F.~Sharpey-Schafer$^{8}$}

\address{$^1$Department of Physics, University of Kashmir, Srinagar,
190 006, India \\
$^2$Department of  Physics, National Institute of Technology,  Srinagar, 190 006, India\\
$^3$Department of Physics, Govt. Degree College Kulgam, 192 231, India\\
$^4$Cluster University Srinagar, Jammu and Kashmir, 190 001, India\\
$^5$Department of Physics, University of Notre Dame, Notre Dame, USA \\
$^6$iThemba LABS, National Research Foundation, P.O. Box 722, Somerset-West 7129, South Africa\\
$^7$Department of Physics, University of Zululand, Private Bag X1001, KwaDlangezwa, 3886, South Africa\\
$^8$University of Western Cape, Department of Physics, P/B X17,
Bellville, ZA-7535, South Africa }

\date{\today}

\begin{abstract}
Excited band structures recently observed in $^{156}$Dy  are investigated
using the microscopic triaxial projected shell model (TPSM) approach and the
quasiparticle random phase approximation (QRPA) based on the rotating mean-field. It is 
demonstrated that new observed excited bands, tracking the
ground-state band, are the $\gamma$-bands based
on the excited two-quasineutron configurations as conjuctured in the
experimental work.
\end{abstract}


\maketitle
\section{Introduction}
A major challenge in nuclear theory is to elucidate the
rich band structures observed in atomic nuclei \cite{BM75}. The
phenomenal progress achieved in
the experimental techniques in recent years has made it possible to probe the nuclear
structure properties at the extremes of spin and isospin. In some heavier nuclei,
more than thirty band structures have been identified and some of these
bands extend up to  angular-momentum of sixty
\cite{JD88,JS87,FG98,JO11}. 
The band structures provide valuable information on the dependence of 
nuclear properties on excitation energy and angular-momentum. For
instance, it is known that pairing correlations are reduced with
angular-momentum due to alignment of protons and neutrons in high-j
intruder orbitals. The  particles or quasiparticles in these high-j and
low-$\Omega$ orbitals have maximum projection along the rotational
axis and demand less collective rotation to generate the
angular-momentum. These quasiparticle configurations then become
energetically favored and cross the ground-state configuration at
a finite angular-momentum, depending on the region. 

In a more recent experimental study \cite{maj14}, the high-spin band
structures in  $^{156}$Dy have
been populated. The most interesting aspect of this investigation is the
observation of the 
$\gamma$-band based on the ground-state up to highest
angular-momentum, I=32 
observed so far. The excited bands that decay to this $\gamma$-band 
have been proposed to  be the $\gamma$-bands based on the
two-quasiparticle configurations. This interesting proposition of the
$\gamma$-bands built on the quasiparticle excitations warrants investigations
using theoretical approaches.

The $\gamma$-band built on the ground-state or vacuum configuration 
was introduced by Bohr and Mottelson \cite{BM75}, who
 interpreted  these $K^\pi =2^+$ bands as
 a dynamic quadrupole deviation of  the nuclear mean-field potential
 from the axial shape. 
In the framework of the Unified Model, it is considered as an
intrinsic excitation which 
when combined with the rotational D-function, restores the angular-momentum. 
 Several approaches  have
been developed to describe  this intrinsic excitation in a microscopic
way that include
quasiparticle phonon \cite{VG81,VG92}, multi-phonon \cite{JL88,MK88,HG88,DG95}, dynamic
deformation \cite{KM84},  the quasiparticle random phase approximation
(QRPA) based on
the rotating mean-field, which describes the $\gamma$-bands based on high-spin yrast levels in a semi-classical way
\cite{SM82a,SM82b,SM83,SM84,KN04,KN07,NK07}.

Recently, the microscopic approach of the triaxial projected shell model (TPSM)
has been developed to describe  the high-spin band structures in
transitional nuclei \cite{JS99}. In this approach, the three-dimensional
projection method is applied to project out the good angular-momentum
states from the triaxial intrinsic states. 
From the symmetry
requirement, the projection from the
self-conjugate vacuum state leads to even-K states with K=0,2, 4,...
The K=0, 2, 4  states represent the main components of the ground-state, $\gamma$- , and $\gamma\gamma$-bands
at low spin.

The TPSM approach  includes multi-quasiparticle
excitations and it is evident from the very construction of the
basis configurations that $\gamma$-
bands can be also built on the quasiparticle configurations.
 This interpretation is similar to the tidal wave approach, which will be applied to the $\gamma$-vibration for the first time in this paper.
It describes the $\gamma$-vibration as a travelling wave, which
corresponds to uniform rotation about the long-axis of  the triaxial
shape. Using the cranking model, the properties of the $\gamma$-vibration are
calculated in a microscopic way without introducing any new parameters.

The recent experimental study of $^{156}$Dy \cite{maj14} provided evidence for
the existence of a $\gamma$-band based on the ground-state band up to very high spins and 
a second  $\gamma$-band, which is proposed to be built on the
s-configuration, which contains 
two rotational aligned i$_{13/2}$ quasineutrons.   
 The purpose
of the present work is to investigate the conjectured $\gamma$-bands,
based on the quasiparticle excitations, using the TPSM
approach as well as in the framework of the traditional approaches based on the 
rotating mean-field. 
 
As the band structures in $^{156}$Dy have been observed up to spin,
I=32, where four-quasiparticle configurations are expected to become
important, four-neutron and four-proton quasiparticle
configurations have been included for the first time in the TPSM basis.
The manuscript is
organized in the following manner :
Section II.A contains  brief description of TPSM approach. Details 
can be found in our earlier publications
\cite{YK00,JY01,YJ02,GH08,JG09}. The results obtained from the TPSM calculations
are presented and compared with the experimental data in section
II.B. In section II.C, nature of $\gamma$-bands is analysed.
Section III.A discusses the structure of the positive parity bands 
in the complementary framework of the conventional cranked shell model and its relation to TPSM interpretation. Section III.B reviews
the application of the quasiparticle random phase approximation to the $\gamma$-vibration in rotating nuclei
from the earlier work in the literature. Section III.C contains the tidal wave study of the $\gamma$-vibration and 
discusses its relation to the TPSM.     
Conclusions are presented in section IV. 

\begin{figure*}[htb]
\vspace{1cm}
\includegraphics[totalheight=13cm]{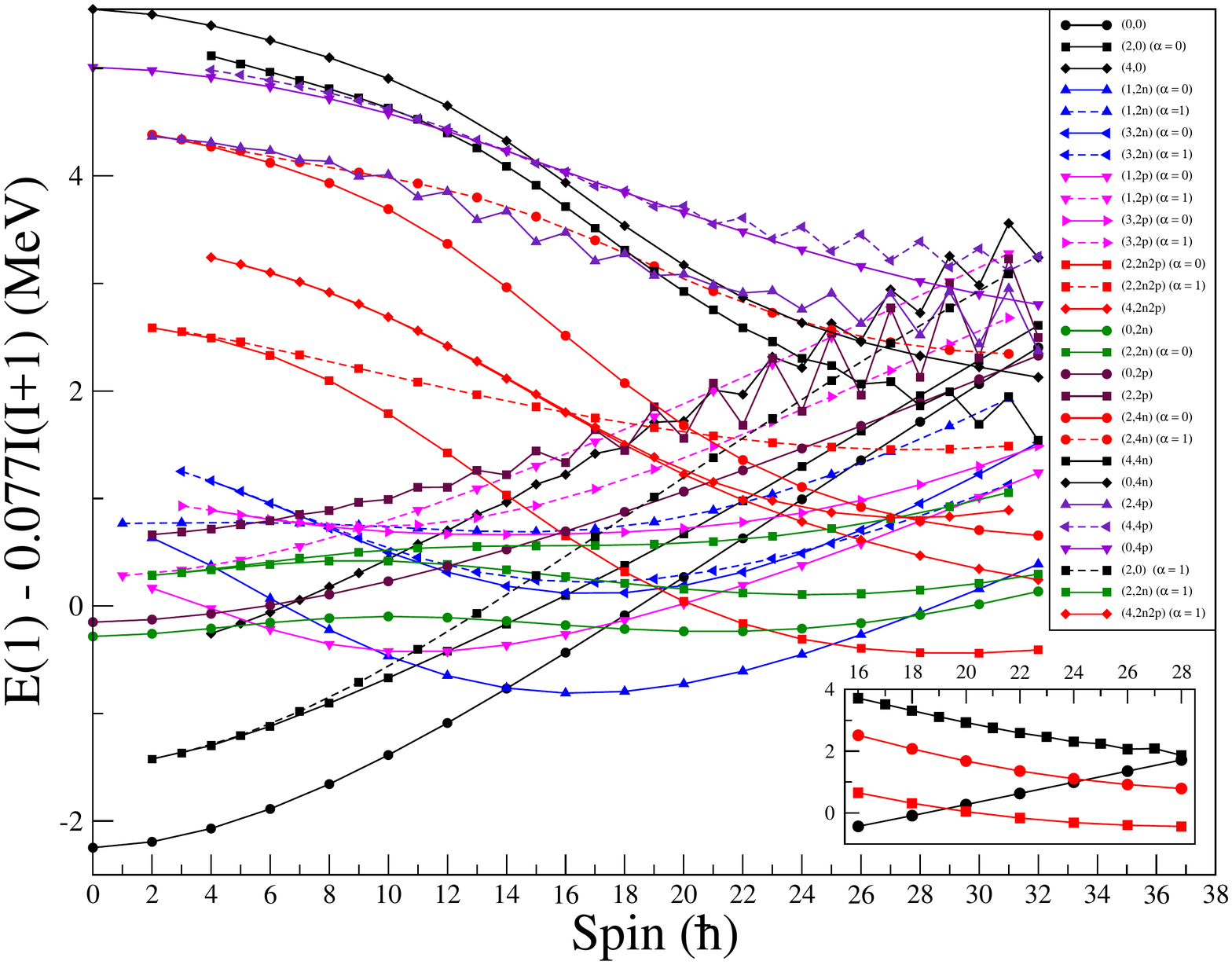}
\caption{(Color online) TPSM projected energies, before band mixing, of
  positive parity states for $^{156}$Dy. Bands are labelled by (K, qp numbers) so that the labels (0,0), (2,0), (4,0),
(1,2n), (3,2n), (1,2p), (3,2p), (0,2n2p)  (2,2n2p) and (4,2n2p) 
correspond to ground,
$\gamma$, 2$\gamma$, two neutron-aligned state, $\gamma$-band on 
this two neutron-aligned state, two proton-aligned state, $\gamma$-band on 
 this two proton-aligned state, two-neutron plus two-proton aligned state,  
$\gamma-$ and $\gamma\gamma-$ band built on this four-quasiparticle
state. The excited K=0 resulting from projection of two-quasineutrons
and two-quasiprotons are denoted by (0,2n) and (0,2p),
respectively. Some bands, having large signature splitting, are
separated into even-spin (labelled as $\alpha=0$)  and odd-spin states
(labelled as $\alpha=1$). The inset depicts the four-quasiparticle
crossings with the ground-state band. The first crossing at I=14 is
due to (1,2n) aligned configuration that forms the s-band
configuration. The second and third crossings at about I=24 and 28 are
due to the aligning 4n configurations with  K=2 and 4, respectively.
}
\label{fig1}
\end{figure*}

\begin{figure*}[htb]
\vspace{1cm}
\includegraphics[totalheight=12cm]{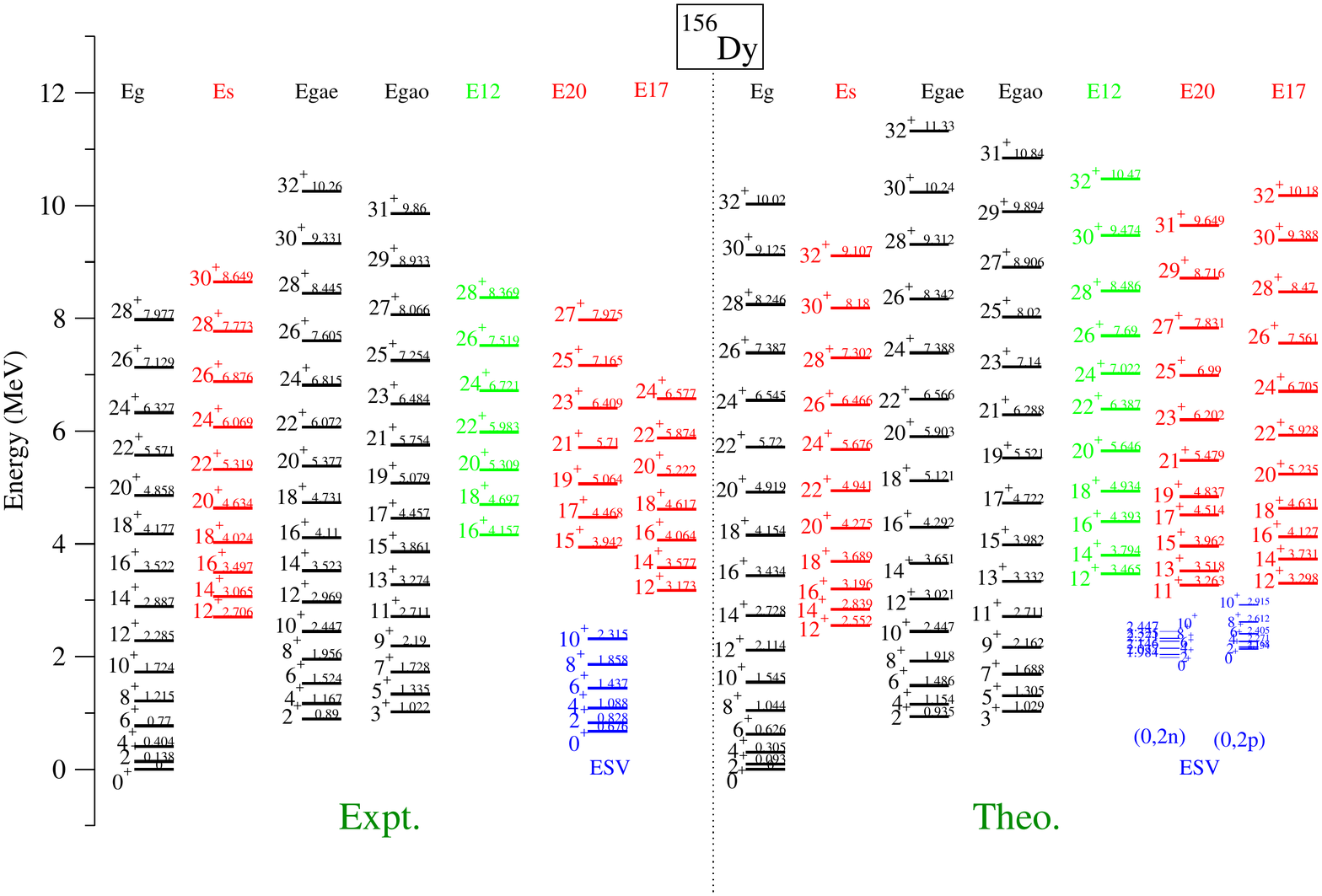}
\caption{(Color online) Comparison of the measured positive parity energy levels of  $^{156}$Dy nucleus \cite{maj14} and the  results of TPSM calculations. 
 }
\label{fig2}
\end{figure*}

\begin{figure*}[htb]
\includegraphics[totalheight=15cm]{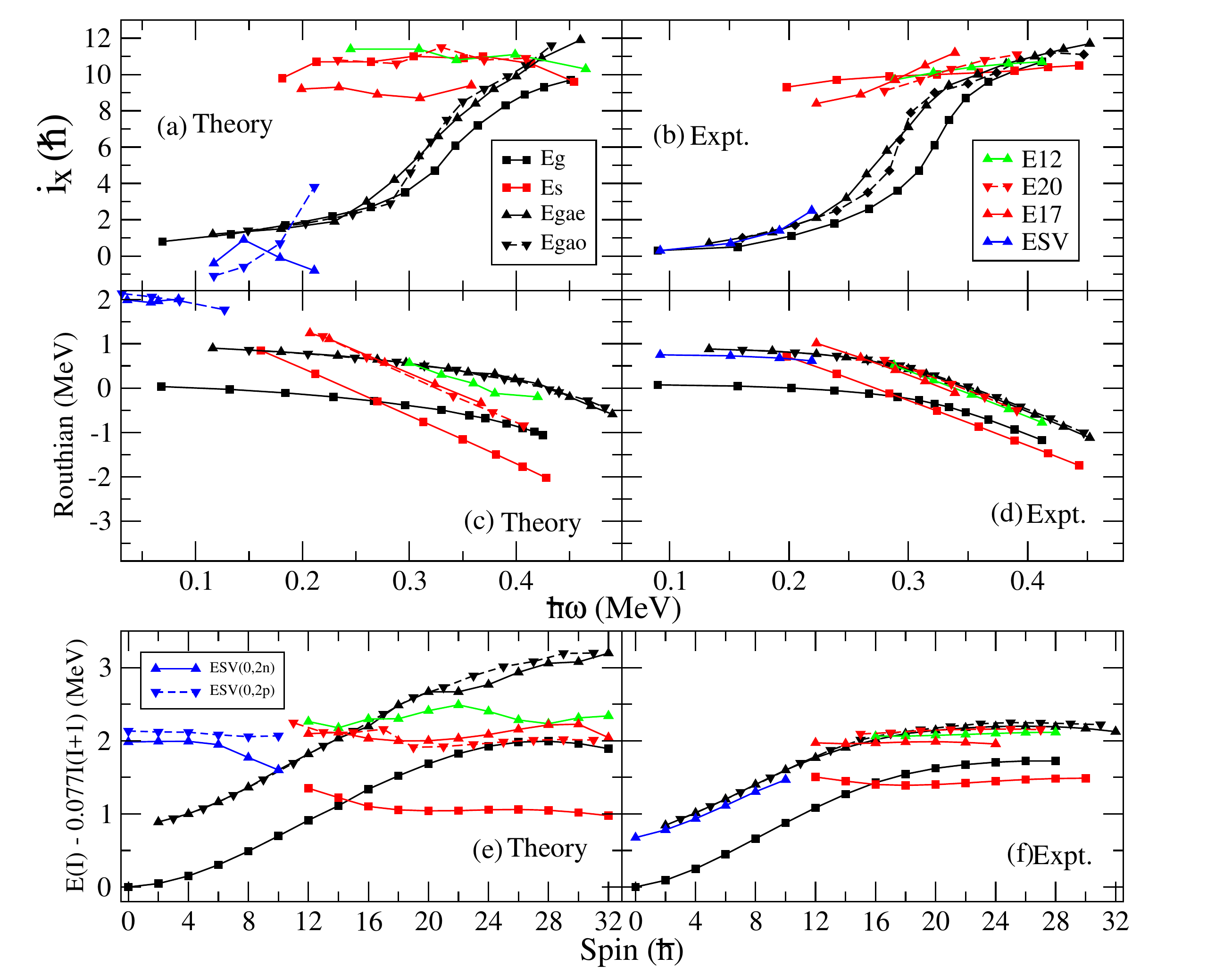}
\caption{(Color online)\label{fig:IxEp_exp_tpsm} Experimental energies
  (right panel)  of the positive parity bands in $^{156}$Dy 
as reported in Fig.~4 of Ref. \cite{maj14}
compared with the TPSM calculations (left panel).
Labeling: Eg - ground-state band, Es - s band, Egae - even-spin $\gamma$-band, Egao - odd-spin $\gamma$-band,
ESV - SV band, E12 - band 12, E17 - band 17, E20 band 20.  
Even-spin states are connected by solid lines odd-spin states by dashed lines. In the upper  panel $I_{ref} =\omega {\cal J}_0+ \omega^3 {\cal J}_1$ is subtracted. In the lower panel 
$E'_{ref}=-\omega^2 {\cal J}_0/2- \omega^4 {\cal J}_1/4$ is subtracted
with ${\cal J}_0=23\hbar^2\mathrm{MeV}^{-1}$
and ${\cal J}_1=90\hbar^4\mathrm{MeV}^{-3}$.
 } 
\end{figure*}


\begin{figure}[htb]
 \centerline{\includegraphics[trim=0cm 0cm 0cm
0cm,width=0.6\textwidth,clip]{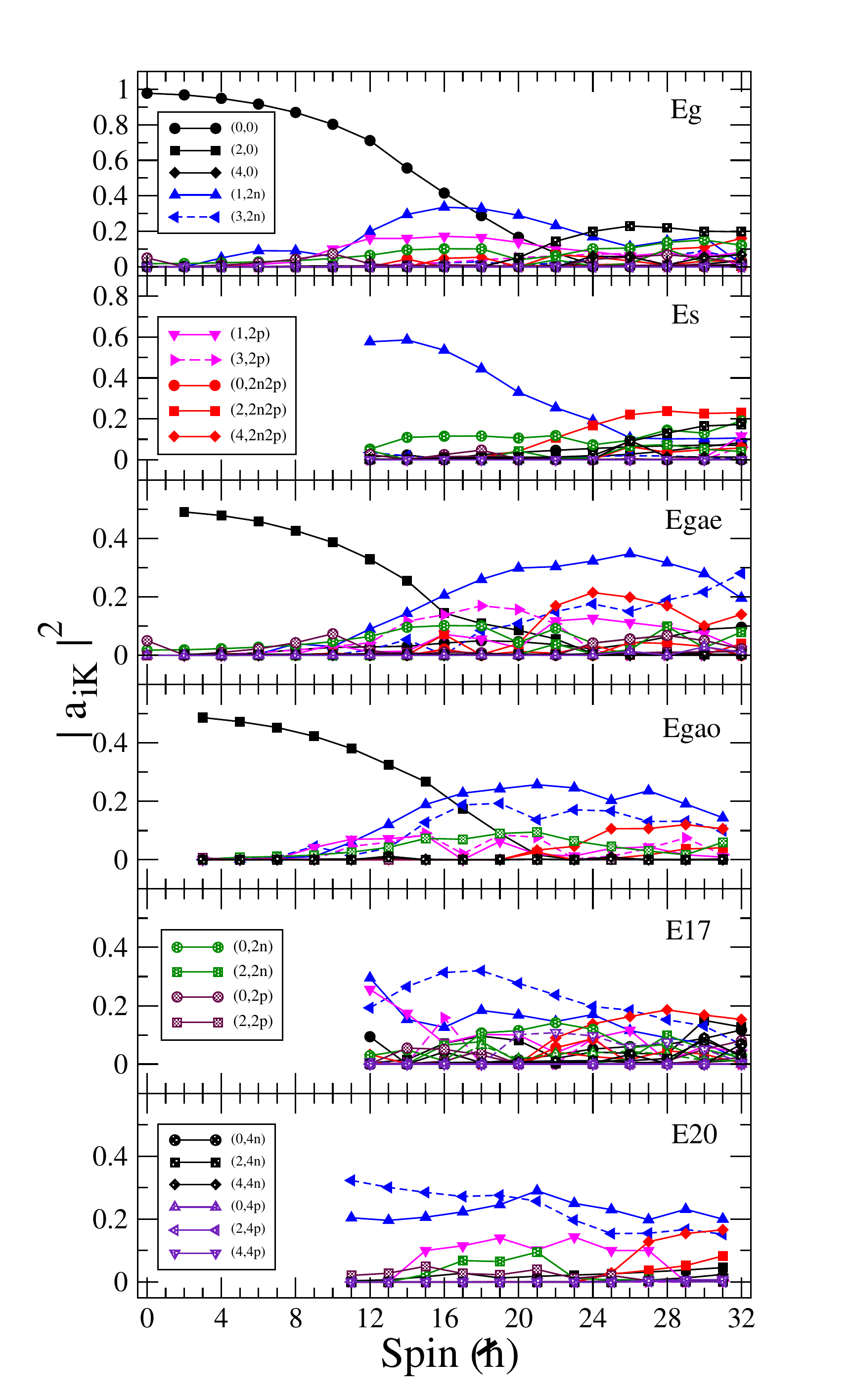}}
\caption{(Color
online) Probability of various projected K-configurations in the
wavefunctions of the band structures after diagonalization are plotted
for the $^{156}$Dy nucleus.} \label{fig4}
\end{figure}


\begin{figure}[htb]
\centerline{\includegraphics[width=\linewidth]{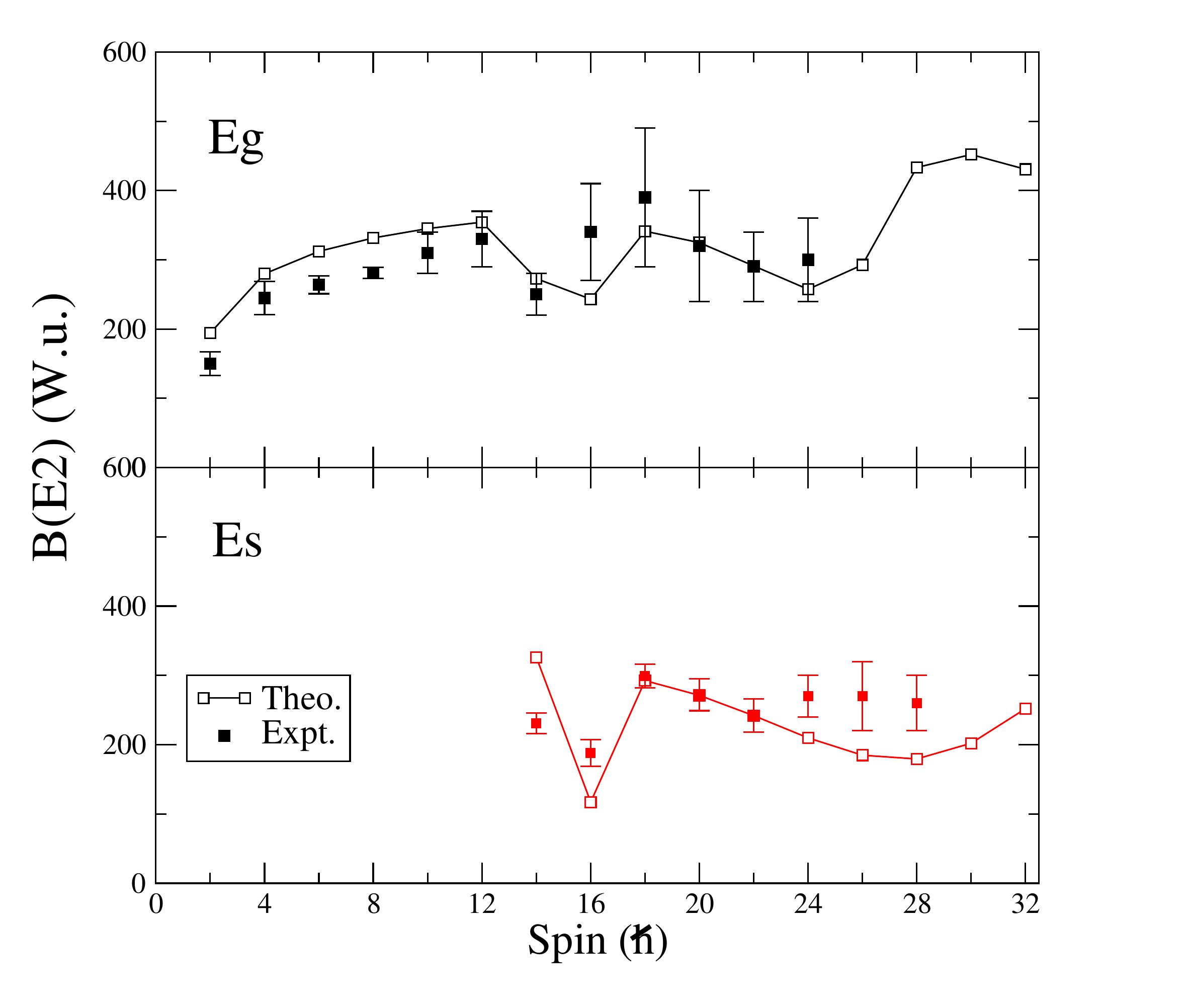}}
\caption{(Color online) Calculated $B(E2)$ transition probabilities
  for the ground-state band and the s-band using the TPSM approach. The
  experimental values have been taken from Refs. 
\cite{CW12,RG86,RP86}.   }\label{fig7pp}
\end{figure}
\begin{figure}[htb]
\centerline{\includegraphics[width=\linewidth,clip]{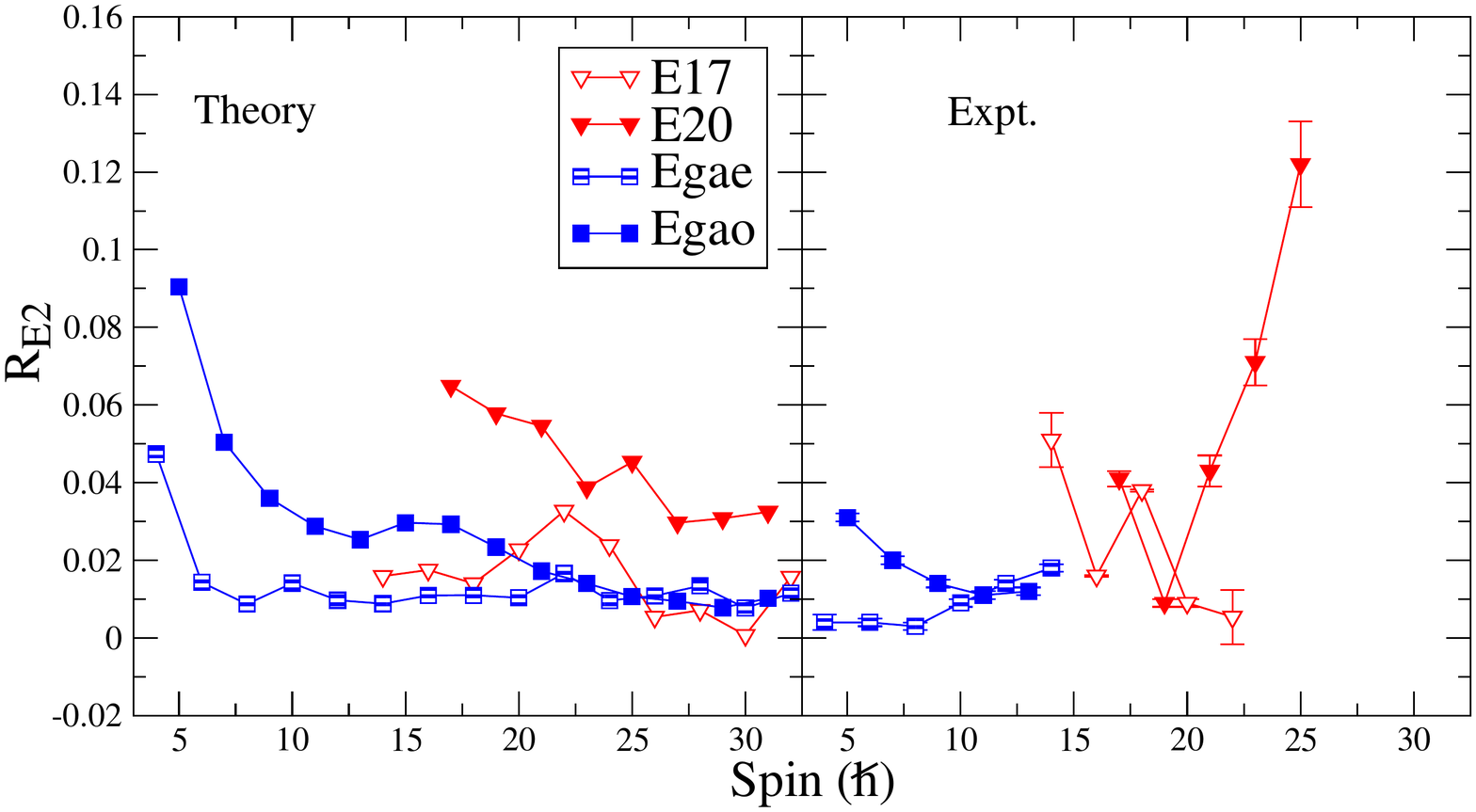}}
\caption{(Color online) Ratios of the $B(E2)$ values for out-of 
band to in-band transitions $R_{E2} = B(E2; out)/B(E2; in)$ for the $\gamma$
decays from the $\gamma$-band to the ground-state band and for the $\gamma$ decays
from bands 17 and 20 to the s-band  }\label{fig8ss}
\end{figure}

\begin{figure}[htb]
 \centerline{\includegraphics[width=\linewidth]{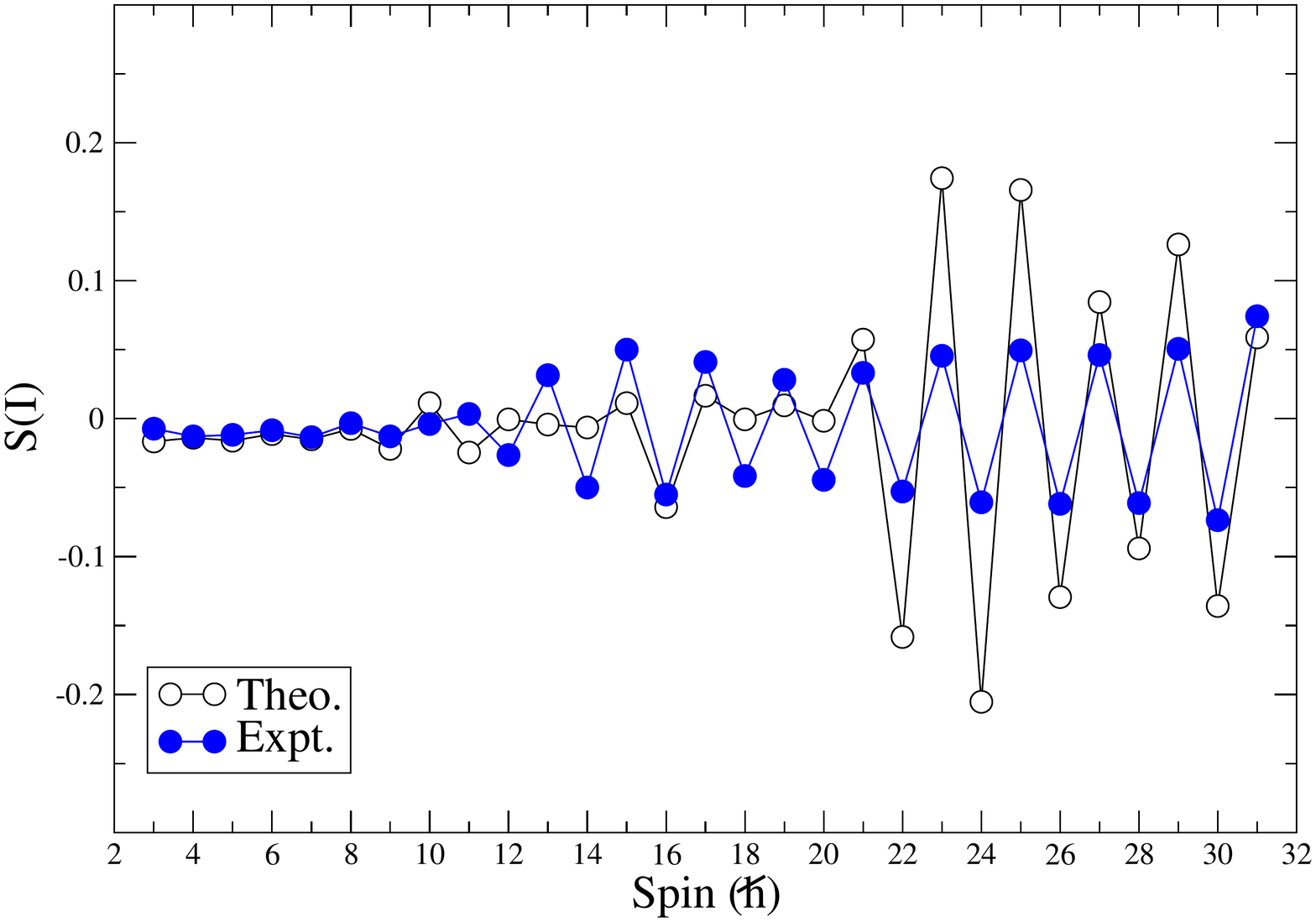}}
\caption{(Color
online) Comparison of observed and TPSM calculated staggering
parameter Eq. (\ref{eq:staggering}) for the
$\gamma$-band in $^{156}$Dy.
} 
\label{fig:staggering}
\end{figure}

\section{TRIAXIAL PROJECTED SHELL MODEL APPROACH}
\subsection{Extension of the model}

For even-even systems, the normal TPSM basis space is comprised of projected 0-qp
or qp-vacuum, 2-proton, 2-neutron and 2-proton + 2-neutron
quasiparticle configurations. In the present investigation of the
band structures in $^{156}$Dy, high-spin states have been
observed up to I=32 and in order to describe the states above I=20 accurately, it
is important to include also 4-neutron and 4-proton quasiparticle
configurations. In the present work, we have extended the TPSM basis 
space to include these four-quaispraticle configuration and the
complete basis space employed in the present work is given by 
\begin{equation}
\begin{array}{r}
\hat P^I_{MK}\ack\Phi\ket;\\
~~\hat P^I_{MK}~a^\dagger_{p_1} a^\dagger_{p_2} \ack\Phi\ket;\\
~~\hat P^I_{MK}~a^\dagger_{n_1} a^\dagger_{n_2} \ack\Phi\ket;\\
~~\hat P^I_{MK}~a^\dagger_{p_1} a^\dagger_{p_2}
a^\dagger_{n_1} a^\dagger_{n_2} \ack\Phi\ket ;\\
~~\hat P^I_{MK}~a^\dagger_{n_1} a^\dagger_{n_2}
a^\dagger_{n_3} a^\dagger_{n_4} \ack\Phi\ket ;\\
~~\hat P^I_{MK}~a^\dagger_{p_1} a^\dagger_{p_2}
a^\dagger_{p_3} a^\dagger_{p_4} \ack\Phi\ket ,
\label{basis}
\end{array}
\end{equation}
where $\ack \Phi\ket$ is the vacuum state and the
three-dimensional angular-momentum projection operator \cite{KY95}
is given by\begin{equation}
\hat P^I_{MK} = {2I+1 \over 8\pi^2} \int d\Omega\,
D^{I}_{MK}(\Omega)\, \hat R(\Omega)~~~.
\label{PD}
\end{equation}
In the above equation, $\hat R(\Omega)$ is the rotational operator in terms of Euler angles.
The adopted projected basis space in Eq.~\ref{basis} is adequate 
enough to describe the high-spin
states observed in $^{156}$Dy up to I=32. $\ack \Phi\ket$
in the TPSM approach is the triaxial quasiparticle vacuum state and
the angular-momentum projection operator in Eq.~\ref{PD} not only
projects out the good angular-momentum, but also states 
having good $K$-values. This is achieved by specifying a value for
$K$ in the rotational matrix, $"D"$, in Eq.~\ref{PD}. 

It is worthwhile to mention that basic strategy in TPSM is similar to that
used in the spherical shell model (SSM) approach except that 
now a deformed basis is employed 
rather than the spherical one. The deformed potential in TPSM 
provides an optimum basis to perform the spectroscopic studies of deformed
heavier nuclei that are presently beyond the reach of the
SSM approach. In the present work, the Wick's theorem is used to evaluate the
matrix elements of rotated many-quasiparticle states. This procedure becomes quite 
involved for more than two-quasiparticles and for identical four-quasiparticle
states, considered in the present development, the number of terms in
Hamiltonian kernel run into thousands. We are presently in the process of implementing
the Pfaffian technique in the TPSM approach to include the 
higher quasiparticle states as has been recently done in the PSM approach \cite{PF1,PF2}.

The projected basis of Eq.~\ref{basis}
is then used to diagonalise the shell model Hamiltonian.
As in our earlier studies, we have employed the pairing plus
quadrupole-quadrupole Hamiltonian \cite{KY95,GH14,JG11,GH144}
\begin{equation}
\hat H = \hat H_0 - {1 \over 2} \chi \sum_\mu \hat Q^\dagger_\mu
\hat Q^{}_\mu - G_M \hat P^\dagger \hat P - G_Q \sum_\mu \hat
P^\dagger_\mu\hat P^{}_\mu .
\label{hamham}
\end{equation}
The $QQ$-force strength $\chi$ is
adjusted such that the physical quadrupole deformation $\epsilon$ is
obtained as a result of the self-consistent mean-field HFB
calculation \cite{KY95}.  The monopole pairing strength, $G_M$, is of
the standard form
\begin{equation}
G_{M} = (G_{1}\mp G_{2}\frac{N-Z}{A})\frac{1}{A} \,(\rm{MeV}),
\label{gmpairing}
\end{equation}
 where $- (+)$ is neutron (proton).  
In the present calculation, we use $G_1=20.12$ and $G_2=13.13$,
which approximately reproduce the observed odd-even mass difference
in this region.  This choice of $G_M$ is appropriate for the
single-particle space employed in the model, where three major
shells are used for each type of nucleon ($N=3,4,5$ for protons and
$N=4,5,6$ for neutrons).  The quadrupole pairing strength $G_Q$ is
assumed to be proportional to $G_M$,  the proportionality
constant being fixed as 0.16.  These interaction strengths are
consistent with those used earlier for the same mass region
\cite{JS99,YK00,KY95}.  

\subsection{COMPARISON WITH EXPERIMENT}
TPSM calculations have been performed for $^{156}$Dy by constructing
the quasiparticle basis space with deformation parameters of
$\epsilon=0.278$  and $\epsilon'=0.105$, which correspond to $\beta=0.29$ and $\gamma=20.6^\circ$ in the standard parametrization.
The axial deformation parameter has been adopted from the earlier
studies \cite{Raman}. The nonaxial
deformation parameter is chosen in such a way that the band-head
of the $\gamma$-band is reproduced.

The angular-momentum projected energies from 0-qp, 2-qp, and 4-qp
configurations, calculated with deformation parameters given above, 
are depicted in Fig.~\ref{fig1}  for $^{156}$Dy. The projection from
0-qp state results into $K=0,2,4,.....$ with no odd-values due to symmetry requirement 
for the vacuum configuration and give rise to ground-, $\gamma$- and
$\gamma\gamma-,.....$bands. The band-head of the
$\gamma$-band is at an excitation energy of  0.98 MeV from the ground-
state and the $\gamma\gamma$-band lies at 
2.13 MeV.  

 Fig.~\ref{fig1} illustrates how the non-rotating quasiparticle basis states  of the TPSM 
 become entangled with increasing angular-momentum.  The ground-state band is
crossed by the signature, $\alpha=0$ branch of $(1,2n)$ band, which 
is a two-quasineutron aligned configuration having K=1, at I=14. 
Further, the $\alpha=0$
component of the two-quasiproton aligned band, (1,2p), with
K=1 also crosses the ground-state band at I=18. There are also crossings in the excited
configuration, for instance, the $\gamma$-band built on the
two-quasineutron configuration (3,2n) having K=3, crosses the normal
$\gamma$-band at I=17. As a matter of fact, the lowest odd-spin states in the spin
range, I=17-25 originate from this configuration. For spin above I=26,
it is noted that four-quasiparticle states (2,2n2p) become yrast. 
In Fig.~\ref{fig1}, the lowest projected K=0 bands, resulting from two-neutron and
two-proton quasiparticle structures, are also plotted. These two bands
are almost degenerate for low-spin states with band-heads at about 2
MeV excitation energy. For high-spin states however the 
K=0 two-quasineutron state becomes favoured. 

The projected bands from four-neutron and four-proton quasiparticle 
configurations are also plotted in Fig.~\ref{fig1} and the band structures
obtained from these states lie at a higher excitation energy compared to
two-neutron+two-proton quasiparticle bands. As is evident from the
figure, these configurations remain higher in energy. However, it is noted
that the $\alpha=0$ component of four-neutron quasiparticle state, having K=2, crosses the
ground-state band at I=24. The projected band from the K=4 component of the
four-neutron state also crosses the ground-state at a higher spin
value. On the other hand, the $\alpha=1$ branch of
this configuration and also the four-proton quasiparticle band
structures lie at higher excitation energies and don't cross the 
ground-state band up to the highest studied spin value. 

In the next stage of the TPSM study, the lowest projected states,
shown in Fig.~\ref{fig1}, and many more ($\sim$ 130 for each angular-momentum)
are employed to diagonalize the shell model Hamiltonian,
Eq.~\ref{hamham}. The energies obtained after diagonalization are
compared with the measured data in Figs.~\ref{fig2}  and \ref{fig:IxEp_exp_tpsm}. 
The bands are
labelled as in the experimental work \cite{maj14}. The association between the calculated and 
experimental bands is discussed below.
In Fig. ~\ref{fig2}, we
provide the exact energies that can be used for further investigations.
Fig. \ref{fig:IxEp_exp_tpsm} demonstrate that the TPSM calculations 
describe well the structure of the yrast region. However, the calculated spectrum is too much spread out in energy.
As discussed in Sec. III.A, this is probably a consequence of the deformation and the pair gaps being kept constant.  

To have a closer comparison between theory and experiment, the excitation energies 
are subtracted by the rotor contribution and the resulting
energies are displayed in Fig.~\ref{fig:IxEp_exp_tpsm} (bottom panels
of e and f). It is evident from the two
figures that TPSM reproduces the experimental data quite reasonably
with the exception for the excited $0^+$ band referred to as the SV
band. The band-head of the experimentally observed band is at 0.7 MeV
excitation and the predicted neutron excited $0^+$ is at about 1.9 MeV.
Theoretically, it is expected to lie at about 2 MeV which is equal to
the excitation energy of the two-quasiparticle states as is evident
from Fig. \ref{fig2}. There are some extra correlations, not included in
the present work, that bring it down to 0.7 MeV and clearly it is 
of considerable interest to investigate this problem in detail. 
It may be of vibrational type, pairing isomer, or a shape coexisting structure. The TPSM
does not include corrections of this type.  

The correspondence between the theoretical and the experimental
band structures plotted in Figs.~\ref{fig2} and \ref{fig:IxEp_exp_tpsm} is made
through the  wavefunction analysis as discussed in the following. For some bands,
it was possible to identify a few lower angular-momentum states, not
observed in the experimental data, through this wavefunction analysis.
The dominant components of the wavefunctions of the band structures
are depicted in Fig.~\ref{fig4}. The ground-state
band, shown in the top panel of this figure, has up to I=14 the largest
component of $(0,0)$ which is the K=0 projected state from the 
triaxial vacuum configuration. It is also evident that the
two-quasiproton component, $(1,2p)$, is building up with spin and
becomes dominant in the spin region, I=18-22. For I=24 and beyond,
four-quasiparticle components are becoming important, in particular,
the K=2 four-neutron configuration. Therefore, this
band is not really the ground-state band as it has dominant
two-quasiparticle and four-quasiparticle configuration for I=18 and
beyond. 
The wavefunction of the s-band, shown in the second panel
of Fig.~\ref{fig4}, has the largest component from the
aligned two-quasineutron state, $(1,2n)$, up to I=24 and beyond this
spin value the four-quasiparticle configurations dominate. 

The largest component in the wavefunction of the even-spin
$\gamma$-band, shown in the third panel of Fig.~\ref{fig4}, up to 
I=14 is $(2,0)$ which is the K=2
projected state from the 0-qp configuration. It is also noted
that the component $(3,2p)$, which is the $\gamma$-band built
on the two-proton aligned band, is quite large in spin regime,
I=14-18 and above I=18 the s-band configuration, $(1,2n)$, becomes 
dominant. For the odd-spin $\gamma$-band, the composition 
of the wavefunction is similar to the even-spin branch. 
The bands 17 and 20, shown in the 5th and 6th panels of Fig.~\ref{fig4},
have interesting structures with dominant component from $(3,2n)$,
which is the $\gamma$-band based on the two-quasineutron
configuration. 

From the above analysis, the emerging picture for the band structures 
labelled as 17 and 20 is quite similar to what has been proposed in the
experimental work \cite{maj14}. The component $(3,2n)$, which is 
$\gamma$-band built on the two-neutron aligned configuration,
is quite dominant in the wavefunctions of the band structures 
labelled as Band 17 and Band 20. These bands are built on the same
intrinsic quasiparticle as that of the s-band of the second panel in
Fig.~\ref{fig4}, except that these are projected with K=3. It is also 
evident from Fig.~\ref{fig4} that these are not purely $\gamma$-bands built on
the two-quasineutron configuration as these also have significant
contributions from other configurations. This is related to the
inter-band mixing, in particular, for high-spin states and also 
due to non-orthogonality of the projected states.

Further, it is interesting to note from Fig.~\ref{fig4} that the Even and Odd
$\gamma$-bands initially have the dominant component  $(2,0)$
as expected, however, for intermediate spin-values, the configuration
$(3,2p)$ becomes important, which is similar to the ground-state band. 
This has to be the case, otherwise the $\gamma$-band would not track 
the ground-state band. This configuration is the $\gamma$-band
built on the two-quasiproton configuration. The ground-state band is crossed by the
normal two-quasiproton configuration and we infer
from Fig.~\ref{fig4} that for the $\gamma$-band, it is $\gamma$-band based on
two-quasiproton state that crosses.

The experimental interpretation of the band structures  as $\gamma$-bands built
on the ground-state and two-quasiparticle configurations is based on the transition probabilities 
between the bands. The transition probabilities have been calculated
in the TPSM approach with the effective charges of  1.5e for protons and
0.5e for neutrons. The
$B(E2)$ values for the transition between the ground-state and the s-bands are depicted
in Fig.~\ref{fig7pp} along with the known experimental values. Two dips in the
$B(E2)$ curve for the ground-state band is due to the crossing
of the two-proton and four-neutron aligned configurations at spin,
I=18$\hbar$ and 24$\hbar$, respectively. A small drop in the $B(E2)$ transitions
of the s-band at I=14$\hbar$ is due to mixing of (2,2n) and
ground-state bands. Further, as in the
experimental work, we have also evaluated the ratios $R_{E2}$ of the
out-of-band $B(E2;out)$ to the in-band $B(E2;in)$ strengths,  where 
the out-of-band transitions connect the $\gamma-$ band 
with the ground-state band and bands 17 and 20 with the s-band. 
[It needs to be added that $B(M1)$ transition probabilities, shown
in Table I, are an order of magnitude smaller than $B(E2)$ transitions
and hence evaluation of branching ratios with $B(E2)$ values only is justified.]
As depicted in Fig.~\ref{fig8ss}, the $R_{E2}$ values
for all the bands are of same order of magnitude, supporting 
the interpretation that band 17 and 20 are the $\gamma$-bands built on
the s-band configuration. The TPSM reproduces fairly well the experimental values.

The wavefunction of Band 12
for spin states of I=16, 18 and 20 are dominated by two-proton aligned
configuration, $(1,2p)$, and above I=24 the wavefunction has 
predominant contribution from four-quasiparticle states. Clearly,
g-factor measurements are needed to probe the intrinsic structures
of these band structures.

To examine further the quasiparticle structures of the observed band
structures, we have analyzed the alignments of the bands as a 
function of the rotational frequency and the
results are presented in Fig.~\ref{fig:IxEp_exp_tpsm} (top panels, a and b). The observed ground-state
band has a gradual increase in the alignment at a rotational frequency of
$\hbar
\omega = 0.3$ MeV. This increase is also noted in the TPSM calculated
alignment, although it is slower than in the
experimental data. This increase can be traced
to the alignment of four-neutrons having K=2. This configuration
crosses the ground-state band at I=24 and becomes the dominant
component in the ground-state band above this spin value 
as is evident from Fig.~\ref{fig4}. Both even- and odd-spin members of the 
$\gamma$-band also show an increase in the
aligned angular-momentum, which is due to the increasing contribution
of $(1,2n)$ component in these band structures.

\subsection{Nature of the $\gamma$-bands}

\begin{table}
\caption{Calculated  $M1\,\,\,(\mu_N^2)$ units  from Egao$\rightarrow$Eg and  E20$\rightarrow$Es of  $^{156}$Dy nucleus. }
\begin{tabular}{|c|c|c|}
\hline 
 Spin ($I^\pi$) & Egao$\rightarrow$Eg    & E20$\rightarrow$Es    \\ \hline
3$^+$      & 0.013       &           \\ 
5$^+$      & 0.011       &           \\
 7$^+$     & 0.010       &            \\
 9$^+$     & 0.008       &           \\
 11$^+$    & 0.003       &           \\
 13$^+$    & 0.002       & 0.124     \\
 15$^+$    & 0.004       & 0.122       \\
 17$^+$    & 0.006       & 0.071      \\
 19$^+$    & 0.008       & 0.029     \\
 21$^+$    & 0.002       & 0.017     \\
 23$^+$    & 0.003       & 0.009     \\
 25$^+$    & 0.005       & 0.005     \\
 27$^+$    & 0.003       & 0.007     \\
 29$^+$    & 0.004       & 0.001     \\
 31$^+$    & 0.001       & 0.002     \\\hline
\end{tabular}
\end{table}

Analyzing the collective Bohr-Hamiltonian results, it has been suggested  that signature splitting of the $\gamma$-band is sensitive to
the nature of $\gamma$ deformation (see e.g the review
\cite{Frauendorf15}). The observed pattern is :  harmonic $\gamma$-vibration about  axial shape - both
signatures degenerate; $\gamma$-independent potential - even spin below odd spin; rigid triaxial potential - odd spin below even spin. 
To quantify the tendency,  the following staggering parameter has been
 introduced : 
 \begin{equation}\label{eq:staggering}
 S(I)= \frac{E(I)-\left(E(I-1)+E(I+1)\right)/2}{E(2^{+}_1)},
 \end{equation}
 which measures the energy of state $I$ relative to the average energy
 of the two neighbours. Fig.~\ref{fig:staggering}
 compares the TPSM results with experiment for the $\gamma$-band on top of the ground-state band.
The staggering  $S(I)$ is quite small at low spin, which is expected for a well established  axial shape. 
Above $I$=10 the staggering increases with the even-spin states below the odd ones.  The 
TPSM calculations reproduce observed staggering pattern, as well as the $R_{E2}= B(E2; out)/B(E2; in)$ values, as shown in Fig. \ref{fig8ss}
 and, therefore, account for the nature of
 the $\gamma$-deformation of $^{156}$Dy nucleus. The even-spin lower,
 which is a signature for $\gamma$-softness, 
 appears to be in conflict with the assumption of the TPSM of a fixed $\gamma$-deformation, which suggests that 
 the criterion based on the Bohr Hamiltonian  does not apply for
 $I>$10 as 
 the TSPM ratios $R_{E2}$ are consistent with experiment.
 Within the TPSM framework,  
 the staggering pattern can be related to the following : for low-even
 spin states the major components are (2,0) and (2,2p). 
The two-neutron aligned configuration, (1,2n),  becomes important
above $I$=10 as is evident from
Fig.~\ref{fig4}, showing the wave function probabilities. For the even-spin members of the $\gamma$-band,  the low-energy even-spin states 
(1,2n) ($\alpha=0$) (cf. Fig. \ref{fig1})  have a large probability.
 For the odd-spin members of the $\gamma$-band  the high-energy odd-spin states 
(1,2n) ($\alpha=1$) (cf. Fig. \ref{fig1})  have a large probability. 
 
 Bands 17 and 20 are assigned to the even- and odd-spin members of  the $\gamma$-band on top of the s-configuration because 
 the basis states (3,2n)   $\alpha=0,1$ are dominant, where   both signatures have nearly the same energy  (cf. Fig. \ref{fig1}).
 As for the $\gamma$-band on top of the ground-state band, both signatures contain a large (1,2n) component of the respective signature,
 which energetically prefers  the even-spin members. 
  
 As seen in Fig. \ref{fig4}, the $\gamma$-bands are distributed over multiple basis states.  As we shall discuss in the next section, the 
 QRPA treatment of the $\gamma$-vibration based on the s-band indicates substantial fragmentation as well. Another reason is the change
 of the quasiparticle structure caused by rotation. Since the TPSM
 uses the quasiparticle configurations in the non-rotational potential as a basis, it describes the modification
 of the quasiparticle structure in the rotating  potential by mixing the non-rotating configurations.

\section{Rotating mean-field interpretation}

\subsection{Cranked Shell Model}

\begin{figure}[t]
\includegraphics[width=\linewidth]{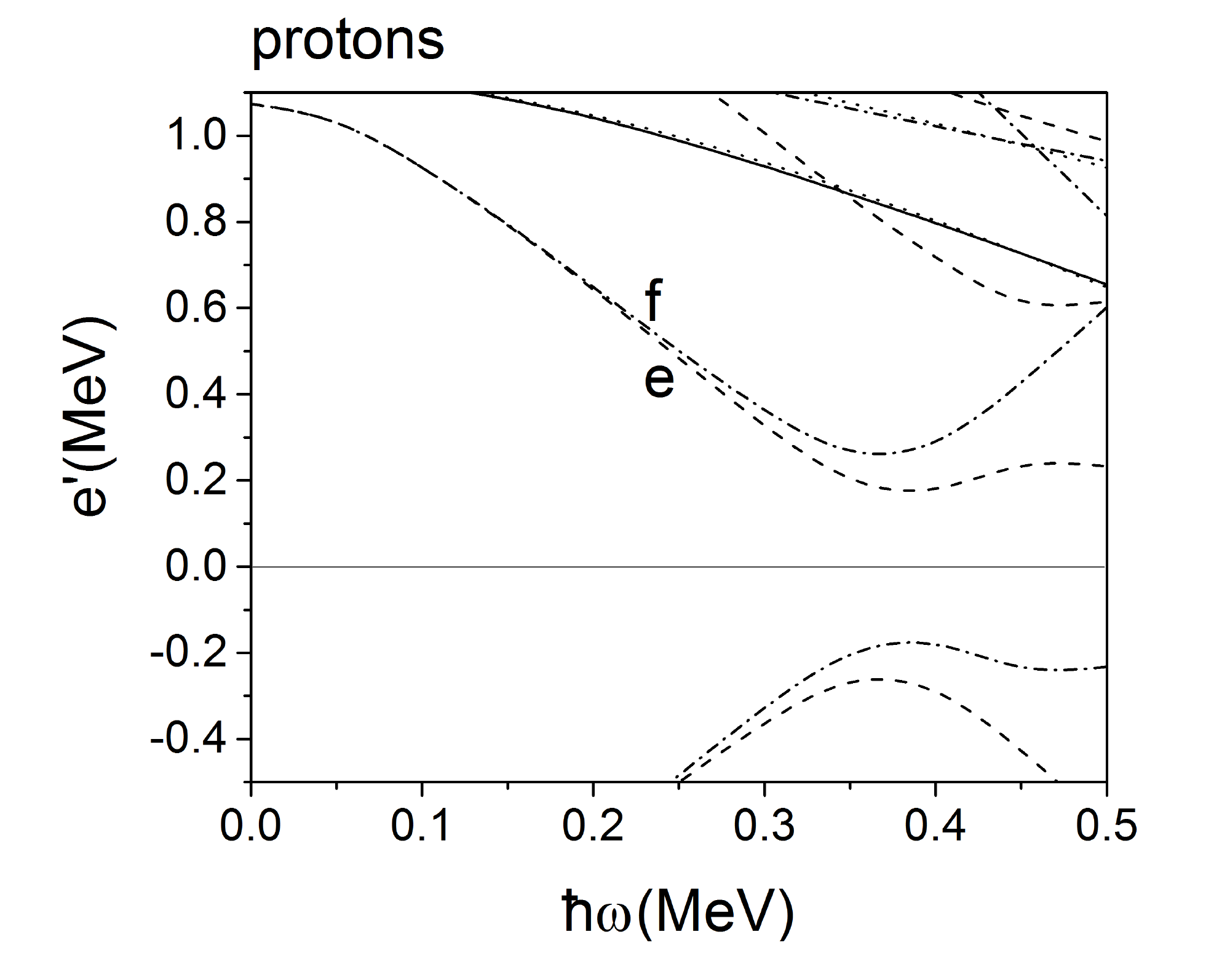}
\includegraphics[width=\linewidth]{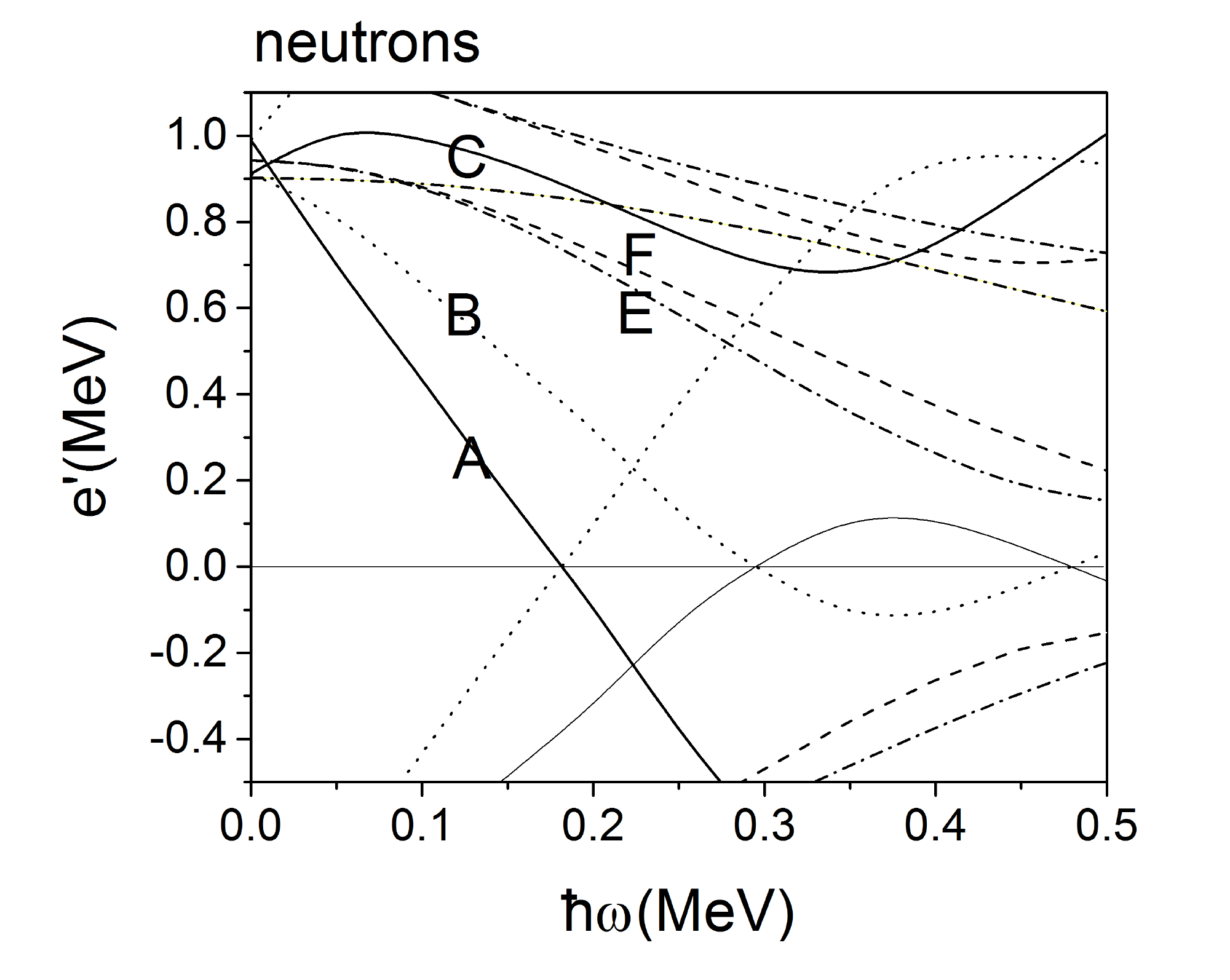}
\caption{ \label{fig:spag} Quasiparticle routhians for $^{156}$Dy.}
\end{figure}

Fig. \ref{fig2} displays the experimental energies and Fig. \ref{fig:IxEp_exp_tpsm} the experimental aligned angular-momenta and routhians
for  the positive parity bands. 
The simplest-possible interpretation is the Cranked Shell Model (CSM) \cite{BF79}, which associates the bands with quasiparticle configurations
in a  potential rotating with frequency $\hbar\omega$ about one of its principal axes. Fig. \ref{fig:spag}  depicts a calculation using the 
axial Nilsson potential 
combined with the  monopole pair field as described in ref.~\cite{BF79}. The  deformation parameters are $\varepsilon=0.26$  and $\gamma=0$, which are close to
the equilibrium deformation found by means of the Cranked Nilsson-Strutinsky method for the relevant spin range \cite{FG98}. The pair gaps are
$\Delta_p=1.0$ MeV and \mbox{$\Delta_n=0.9$ MeV}.   The ground-state band is represented by the vacuum denoted by ``0''.
Below $\hbar \omega$= 0.15 MeV,  it corresponds to all negative-energy quasiparticle routhians occupied. Above it corresponds 
to the diabatic continuation of this configuration. The other configurations are denoted by the occupied routhians labelled by the letters
(the reflected-through-0 routhians are unoccupied). 

\begin{figure}[t]
\includegraphics[width=\linewidth]{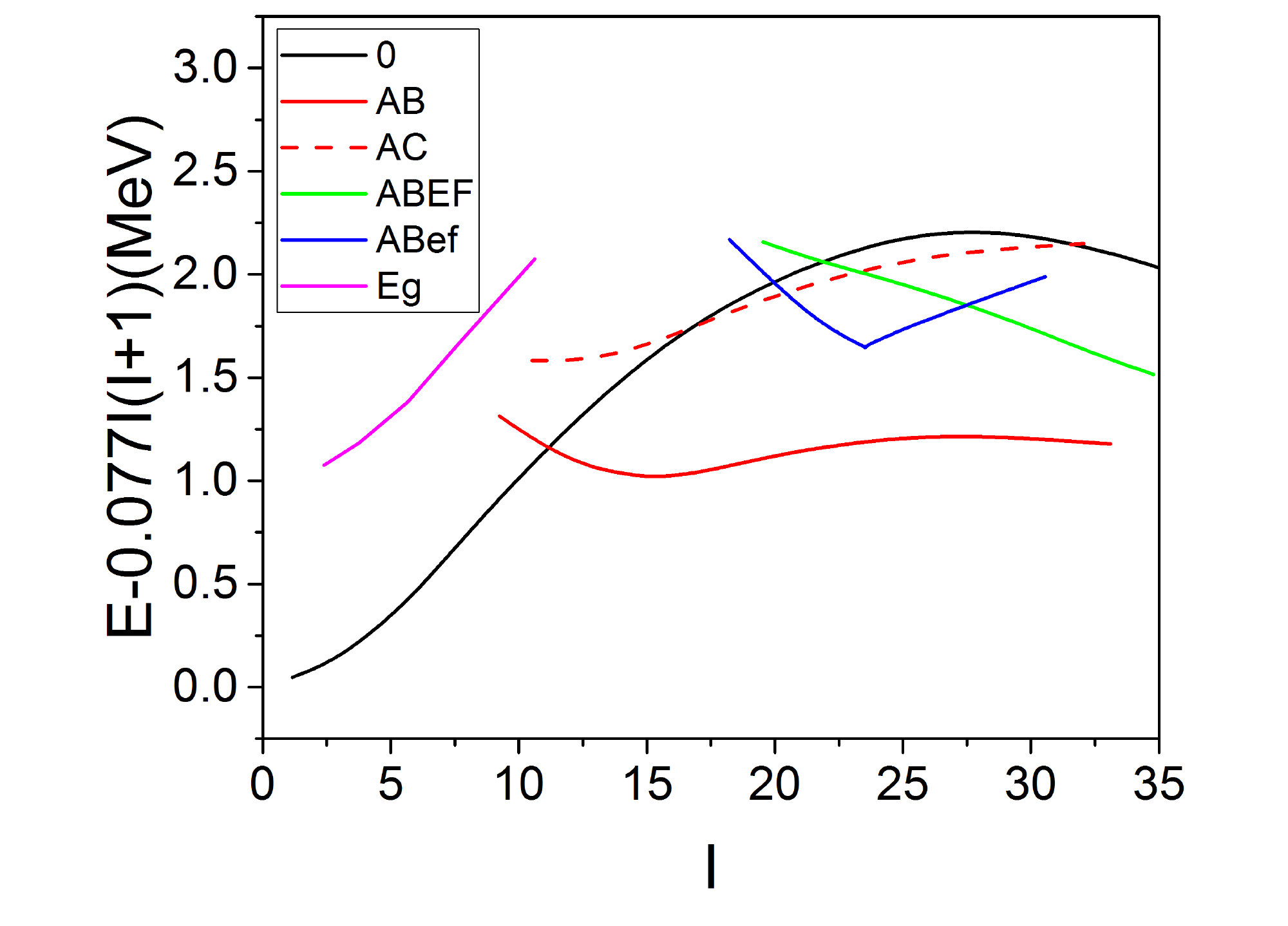}
\caption{\label{fig:Etac} Energies of the lowest positive parity bands in $^{156}$Dy calculated by means
of the cranking model. The bands are labeled by the occupied quasiparticles in Fig.  \ref{fig:spag}.
Eg is the $\gamma$-band calculated by means of the tidal wave approach.
Even-spin states are connected by straight lines odd-spin states by dashed lines.
} 
\end{figure}

\begin{figure}[t]
\includegraphics[width=\linewidth]{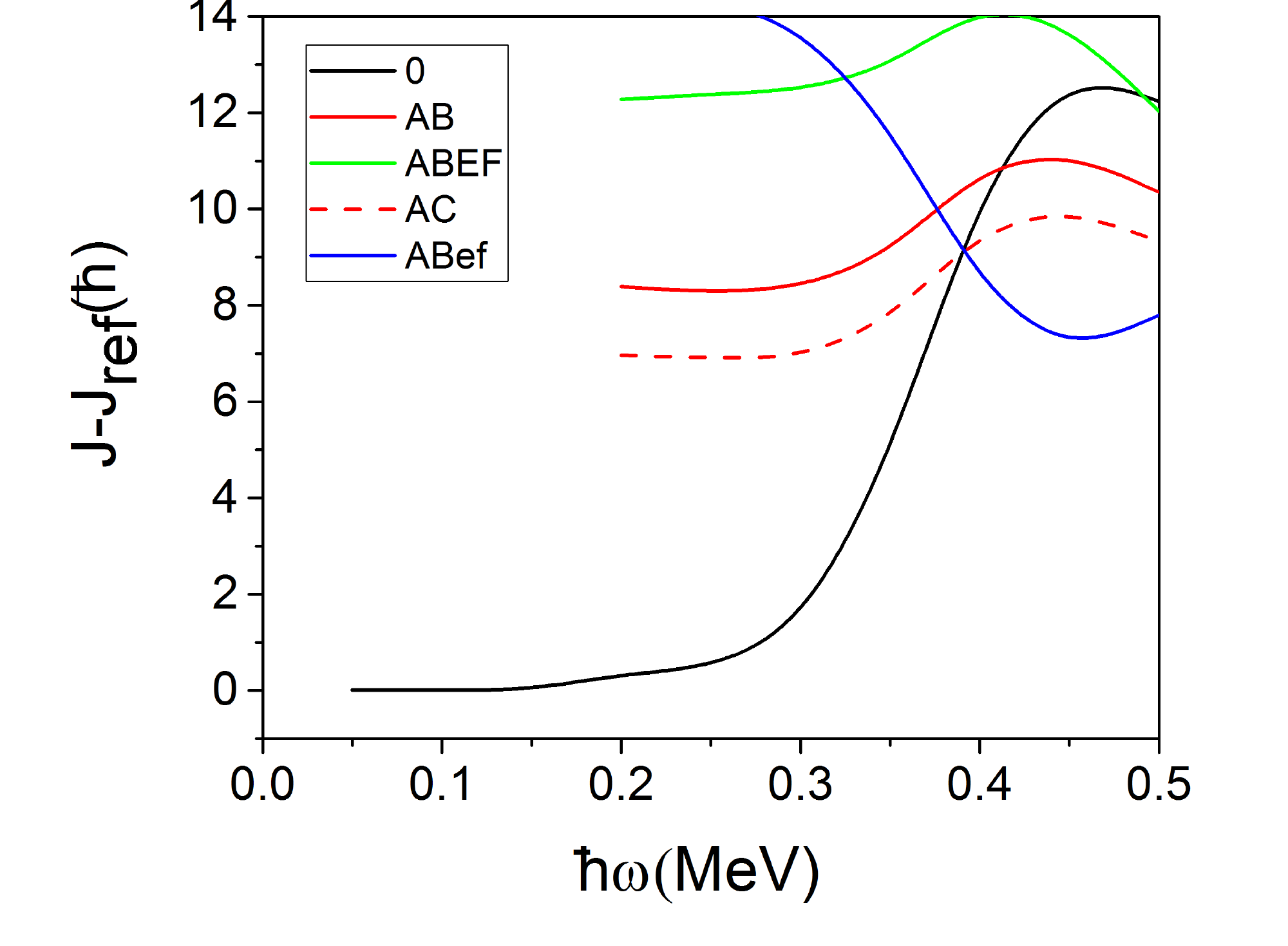}
\includegraphics[width=\linewidth]{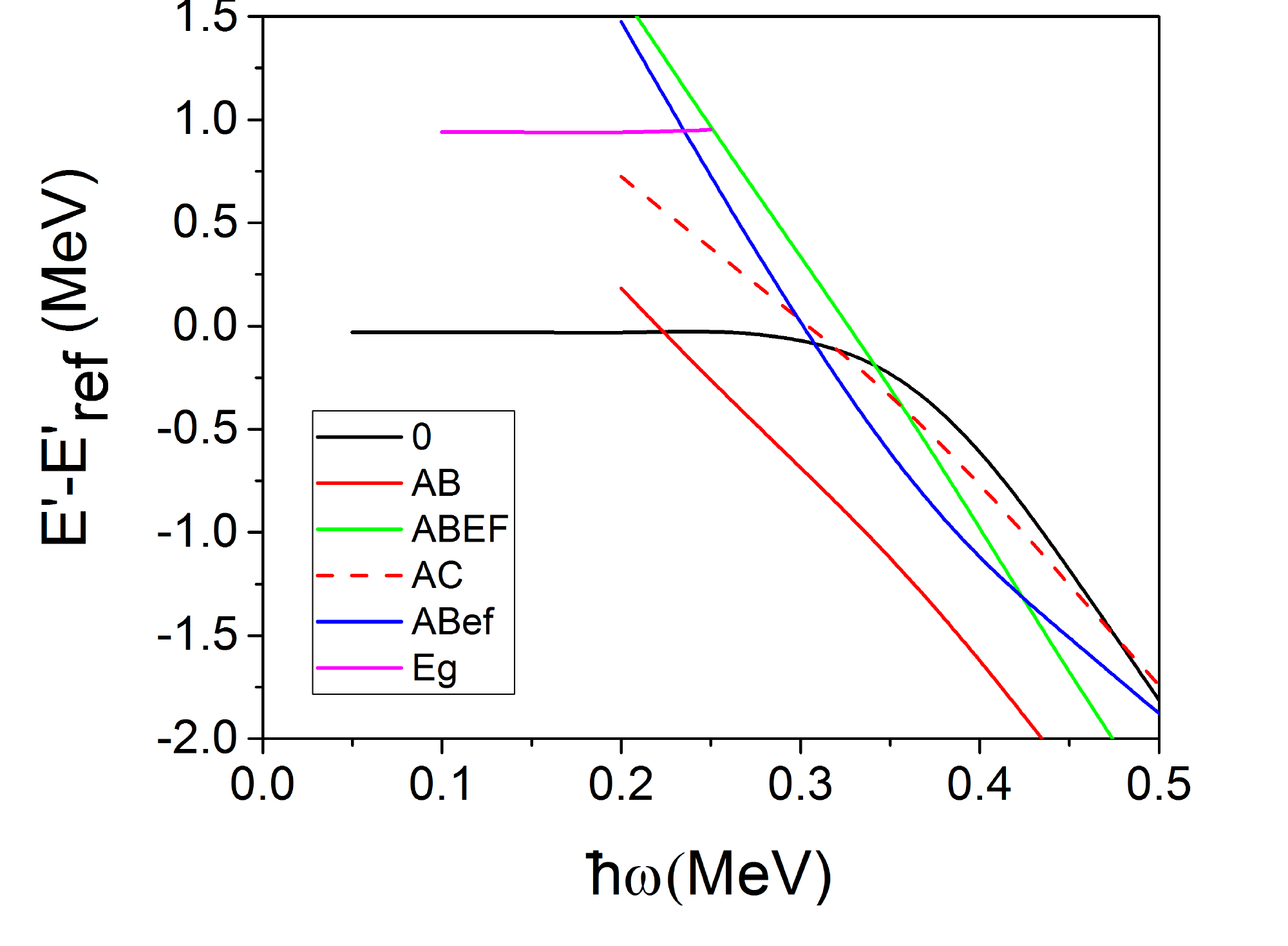}
\caption{\label{fig:IEptac} 
Upper panel: Angular-momentum expectation values calculated by means
of the cranking model relative to the reference. 
Lower panel: Routhians relative to the reference. Reference as in Fig. \ref{fig:IxEp_exp_tpsm}. } 
\end{figure}

Figs. \ref{fig:Etac} and \ref{fig:IEptac} show the lowest positive parity bands labelled by their quasiparticle configurations.
Comparing the CSM calculation with the experiment shown in Fig. \ref{fig2} and \ref{fig:IxEp_exp_tpsm}, 
one notices that the crossing between the g- band (configuration  0) and the s-bands (AB) is qualitatively described. The CSM crossing frequency
of $\hbar \omega =0.23$ MeV  
underestimates the experimental value of $\hbar \omega =0.28$ MeV,
which is a well known deficiency of the CSM. The aligned 
angular-momenta are of right magnitude.
The distance between the two bands after their crossing is overestimated (about 1 MeV in experiment vs. about 2 MeV in the calculation).
We attribute this to the assumption of a fixed deformation and fixed pairing gaps. The Cranked Nilsson-Strutinsky calculations of Ref. \cite{FG98}, which assume no pairing and 
optimize the deformation for each configuration, obtain  a more compressed spectrum, close to experiment.  The CSM gives two even-spin (ABEF and ABef) and one 
odd-spin (AC) bands between the s- and the g-bands, where Ref. \cite{maj14} suggests the experimental location of
 the two signatures of the $\gamma$-band on top of the s-band. The close neighborhood  may cause 
 fragmentation of the collective strength    among these two-quasiparticle  (relative to AB) excitations.

\subsection{Quasiparticle Random Phase Approximation}
Quasiparticle Random Phase Approximation is the standard extension of  the 
 mean-field approximation to describe collective excitations. 
In order to describe  the $\gamma$-vibrations in rotational nuclei, the authors of  Refs. 
\cite{SM82a,SM82b,SM83,SM84,KN04,KN07,NK07}  applied QRPA  to the rotating mean-field  
and the pairing + quadrupole quadrupole model Hamiltonian, Eq.~\ref{hamham}, below with $G_Q=0$. Since the rotating
mean-field conserves signature, the QRPA leads to two independent sets
of equations, one for even and the other one for odd spins.   

\begin{figure}[t]
\includegraphics[width=\linewidth]{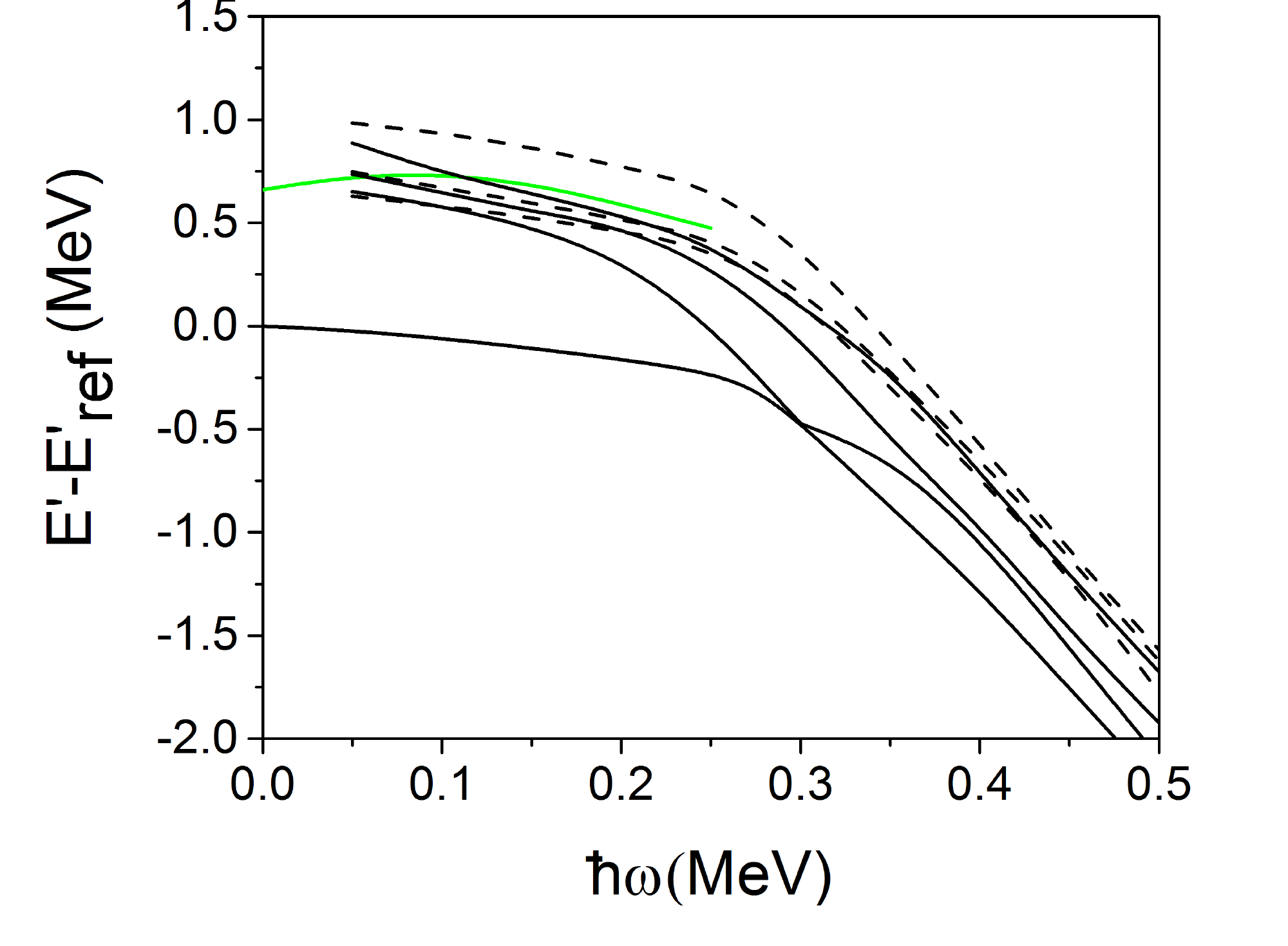}
\caption{\label{fig:qrpa} Routhians  of the lowest positive parity bands in $^{156}$Dy calculated by means
of the QRPA extension of the rotating mean-field derived from the pairing + quadrupole quadrupole
model Hamiltonian. Even-spin states are connected by straight lines odd-spin states by dashed lines.
Prepared using the results of Refs.  \cite{KN04,KN07,NK07}. Reference as in Fig. \ref{fig:IxEp_exp_tpsm}.
   } 
\end{figure}

 Fig. \ref{fig:qrpa}
 shows the QRPA results of Refs.~\cite{KN04,KN07,NK07}. The authors used the  self-consistency requirement
 for a harmonic oscillator potential and volume conservation introduced by Bohr and Mottelson  \cite{BM75}  to determine the quadrupole deformation and to fix the 
 coupling constant   $\chi $.  
The calculations reproduce  very well the experimental energies of the odd-spin branch of the $\gamma$-band on top of the g- band.
The calculated branching ratios $R_{E2}=B(E2, I\rightarrow
I-1:out)/B(E2, I\rightarrow I-2:in)\sim 0.02$ between 
the transitions out-of and within 
the $\gamma$-band account for the measured ratios \cite{maj14} (see Fig. \ref{fig8ss}).  
 Above  the crossing between the g- and s- bands, the QRPA
 solutions are based on the s-band, which is yrast. The three lowest odd-spin solutions are found close together.
 This suggests that the collective vibrational strength 
 may be fragmented over these states instead of being concentrated in one individual  state.   The authors quote only the ratio
 $B(E2, I\rightarrow I-1:out)/B(E2, I\rightarrow I-2:in)\sim 0.04$    for the lowest band.
  The value is consistent with the widely scattered experimental ratios of 
 band 20 (see Fig. \ref{fig8ss}), who's excitation energy with respect to the s-band of about 0.75 MeV 
  is somewhat larger than the QRPA value of about 0.60 MeV. The lowest three
 even-spin QRPA solutions are close together as well, which may lead to fragmentation. The lowest  
 solution lies about 0.2 MeV above yrast, which is substantially
 lower than the position of band 17, the experimental candidate for the even-spin $\gamma$-band. 
  The   calculated ratio  $B(E2, I\rightarrow I-2:out)/B(E2, I\rightarrow I-2:in)\sim 0.01$  is to be     
  compared with the experimental ratio $\sim 0.2$.

The QRPA approach becomes unreliable in the vicinity of the crossing
of the g- and s- bands. The reason is that 
g- and s- configurations have different deformations (0 axial and AB triaxial).
The cranking model produces a mixing between the configurations, which  
 makes the even-spin QRPA energy approaching zero, which is an artifact. 
 The mixing falsifies  the energies of the lowest QRPA even-spin solution already away from 
 the crossing. In contrast to experiment, the even-spin solution (not shown) was  found below the odd-spin sequence.
 The low energy of the even-spin QRPA solutions on top of the s- configuration may be an artifact as well.
  The authors of Refs. \cite{SM82a,SM82b,SM83,SM84}
 avoided these problems by removing the mixing between the g- and s- configurations. 
  For the studied nucleus $^{164}$Er, both the even-spin and odd-spin
 QRPA solutions are stable in the crossing region.

As seen in Fig. \ref{fig:qrpa}, The QRPA calculations  \cite{NK07}
reproduce the experimental energies of the SV band very well.
The structure of the QRPA solution is not analyzed in detail. By construction, it has to be a combination of a $\beta$-type and a pairing vibration.

\subsection{Tidal Wave Approach}

The tidal wave concept has been developed to describe the yrast sequence of near-spherical and transitional nuclei \cite{tidal,102pd}.
It can be directly applied to the $\gamma$-degree of freedom.
The $\gamma$-vibration carries $2\hbar$ of angular-momentum along the symmetry axis.  Classically, the sequence 
$n$~=~1, 2, 3, ... of aligned $\gamma$ phonons  is represented by a wave travelling around  the symmetry axis 3 with the
angular velocity $\hbar \omega_\gamma=E_\gamma/2$ carrying the angular-momentum $J_3=K\hbar=2n\hbar$.
  In the co-rotating frame, the wave is represented 
by a constant $\gamma$-deformation, which increases $\propto \sqrt{n}$. For a travelling wave  the angular-momentum and energy
increase by increasing the amplitude while the rotational frequency stays constant.    
For a triaxial rotor the states above the yrast sequence  are also generated by adding quanta of 
angular-momentum along the axis with the smallest moment of inertia. The difference to the tidal wave 
is that the angular-momentum and energy
increase by increasing the rotational frequency  while  the deformation stays constant.  Obviously, the concept also comprises
the  intermediate cases of an anharmonic $\gamma$-vibration and a soft triaxial rotor. 

It is important to realize that it is not possible to distinguish between a triaxial rotor and a harmonic
$\gamma$-vibration, as long as one considers only the first excited (K=2) band. In both cases an angular-momentum of $J_3= 2\hbar$
is generated by a $\gamma$-deformed shape rotating about the
3-axis. To differentiate one has to 
take the next (K=4) band into consideration. 
The $K$ - dependence of $\omega_\gamma$ is indicated by the level distance $\hbar \omega_\gamma=(E_\gamma(K)-E_\gamma(K-2))/2$.  
In the case of a vibration,  $\omega_\gamma$ is the same  for the $K=2$ and $K=4$ bands and  deformation parameter $\gamma(K=4)= \sqrt{2}\gamma(K=2)$,
which means $E_\gamma(K=4)=2E_\gamma(K=2)$.
In the case of the rigid triaxial rotor,   $\gamma(K=4)= \gamma (K=2)$ and $\omega_\gamma(K=4)=2 \omega_\gamma(K=2)$, which means $E_\gamma(K=4)=4E_\gamma(K=2)$.
This explains why  the TPSM approach \cite{JS99} which operates with a fixed $\gamma$ deformation describes the first excitations of  $\gamma$ vibrational
type in $^{156}$Dy, which has an axial shape at the moderate spins of interest in this communication.

\begin{figure}[t]
\includegraphics[width=8cm]{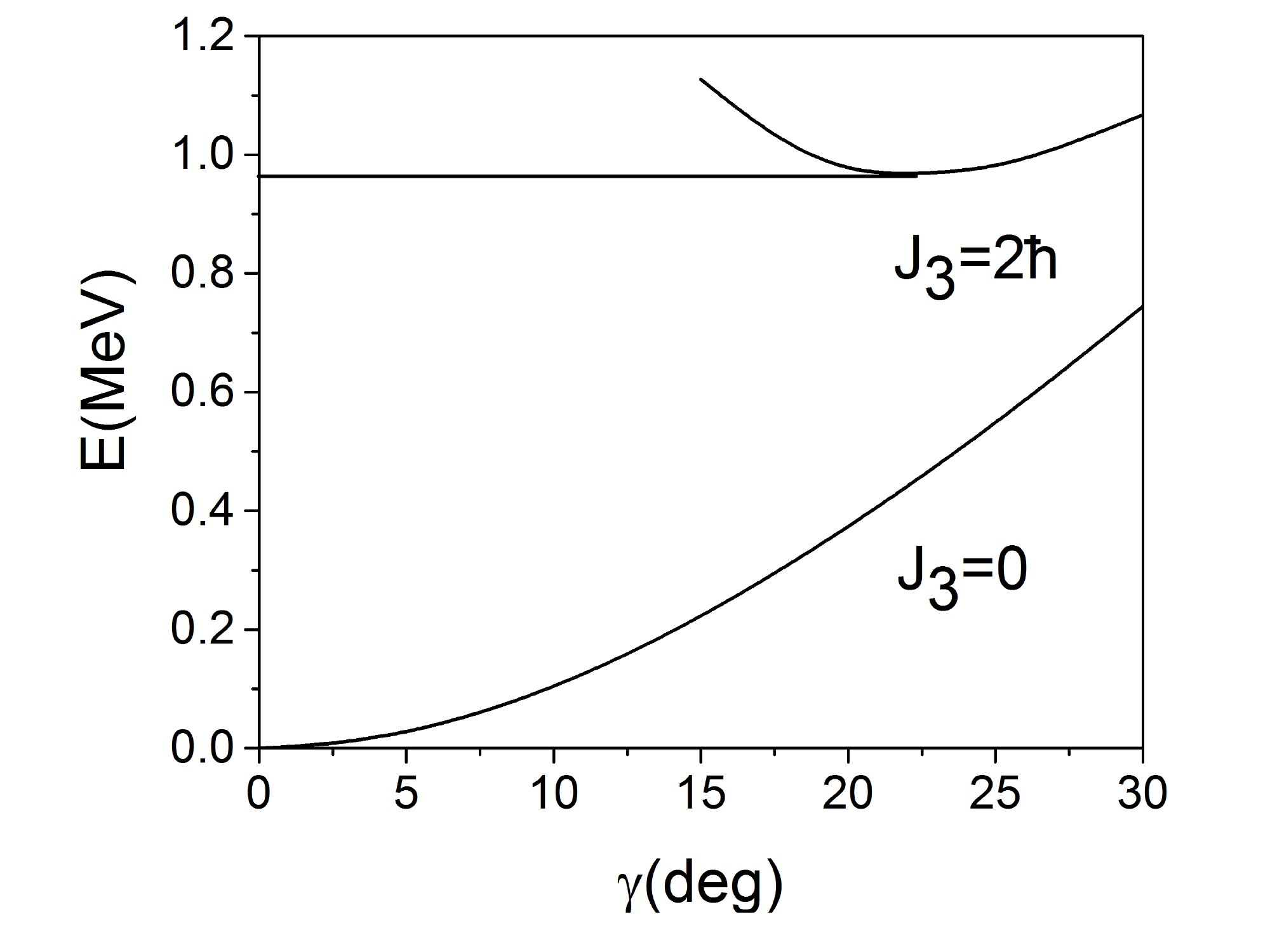}
\caption{\label{fig:tidal} Energy of the $J_3=0$ and $J_3=2\hbar$ calculated by the 
shell correction version of the cranking model.    } 
\end{figure}

For the first time, we apply the Tidal Wave concept to the $\gamma$ vibration in a quantitative way. The travelling wave is described by cranking
the triaxial potential about the long-axis with the frequency $\omega_\gamma$. The total angular-momentum 
$J_3( \omega_\gamma,\gamma)=\langle 0\vert j_3\vert 0 \rangle$ 
and the total routhian  $E'( \omega_\gamma,\gamma)$ are  calculated by means of the 
shell correction version of the cranking model as described in Ref. \cite{sctac}. 
The  total energy $E(J_3=2,\gamma)=  E'( \omega_\gamma,\gamma)+\omega_\gamma J_3( \omega_\gamma,\gamma)$
is minimized with respect to $\gamma$, where $\omega_\gamma$ is fixed by requiring $J_3( \omega_\gamma,\gamma)=2\hbar$.

The calculation is carried out for the vacuum configuration $\vert 0, \omega_\gamma\rangle$, which corresponds to all negative energy quasiparticle routhians occupied. 
Like for rotation about the short-axes 1, 2  shown in Fig. \ref{fig:spag}, the continuation of the configuration  to frequencies $\hbar \omega_\gamma>0.3$ MeV
leads into the region where the negative- and positive-energy quasiparticle encounter. The cranking model generates an unphysical mixing of the 0 configuration
with high-j configurations (AB for example). In order to remove the mixing, the low-frequency routhians are extrapolated using fourth-oder polynomials, which corresponds to
third-order perturbation theory with respect to $\omega_\gamma$. The Harris parametrization :
\begin{equation}
J_3=\omega_\gamma {\cal J}_0+\omega_\gamma^3 {\cal J}_1,~~~
E'=E_0-\frac{\omega_\gamma}{2} {\cal J}_0-\frac{\omega_\gamma^4}{4} {\cal J}_1~~~~,
\end{equation}
is fitted to the cranking values in the range $0\leq \hbar\omega_\gamma \leq 0.1$ MeV. The extrapolation very well reproduces the cranking calculations up to the region
where the mixing with the high-j configurations sets in. 

Fig. \ref{fig:tidal} shows that the energy with $J_3=0$ is minimal for $\gamma=0$, which indicates that the $\gamma$-band has the character of a  tidal wave.
For $J_3=2 \hbar$ the minimum lies at $\gamma=22.5^\circ$, which is the amplitude of the wave travelling about the symmetry axis.  The energy of the minimum 
$E_\gamma=0.97$~MeV is somewhat larger than the experimental energy of the 
$\gamma$-band head of $E(2^+_2)=0.890$~MeV.  The energy $2\hbar\omega_\gamma$=0.91~MeV corresponding to angular velocity of the wave  at the minimum of 
$\omega_\gamma=0.456$MeV$/\hbar$, remarkably well  reproduces  the experimental $2^+_2$ energy.  Up to spin I=10 the coupling of the $\gamma$-band to
other bands is relatively weak. Accordingly, the band-head energy is added to the g-band in Fig. \ref{fig:Etac}.   
Using the  semi-classical expression given in Ref. \cite{sctac}, the cranking calculations provide the   $B(E2,\omega)_{TAC}$ values.
 A rough estimate of the reduced transition probability
 $B(E2,2^+_2\rightarrow 0^+_1)$  is given by the semi-classical expression for  $J_3=1\hbar$,   
 because the latter is proportional to $J_3$. The calculated ratio 
 \begin{equation}
 \frac{B(E2,2^+_2\rightarrow 0^+_1)}{B(E2,2^+_1\rightarrow 0^+_1)}=\frac{B(E2,\omega_3(J_3=1\hbar)_{TAC}}{B(E2,\omega_1(J_1=2\hbar)_{TAC}}
 \end{equation}  
of 0.043 is close to the experimental ratio of 0.048 \cite{ensdf}. 
. 

The tidal wave approach reproduces the properties of $\gamma$-band
head remarkably well without introducing any new parameters. 
This approach becomes equivalent to the QRPA approach when the two approaches are applied to the  pairing + quadrupole quadrupole
Hamiltonian and the linear extrapolation is used for $J_3(\omega_3) $ instead of the third order Harris expression.  
The TPSM, also describes the $\gamma$-excitations by assuming a constant $\gamma$-deformation. Instead of the
semi-classical cranking model it generates the travelling wave by means of  quantal angular-momentum projection.  The $\gamma$-deformation 
is considered as a parameter that  is adjusted to reproduce the excitation energy of the $\gamma$-band on top of the ground-state band. 
The value of $\gamma=20.6^\circ$ is close  to $22.5^\circ$ calculated  by the tidal wave approach.  
As discussed above this does not mean that $^{156}$Dy has a rigid triaxial shape. As long as only the first excited $\gamma$-band is of interest 
there is no way to decide whether the triaxiality is static or dynamic.  The wavelength of the first excitation is too large to resolve details of the
potential in the $\gamma$-degree of freedom that separate a harmonic travelling wave from  a rigid rotor.   


\section{CONCLUSIONS}
The recently observed band structures in $^{156}$Dy
have been interpreted in the framework of the TPSM approach 
and QRPA based on the rotating mean-field. 
The $\gamma$-band built on the ground-state band is well reproduced by the TPSM, in particular,
the staggering pattern above $I=10$, which within the collective model indicates the even-$I$-low pattern,
corresponding to the $\gamma$-softness limit.  
 The TPSM analysis strongly supports  the interpretation  proposed in the experimental work 
 that two excited bands are the $\gamma$-bands based
on the neutron s-band. This is the first detailed confirmation of
$\gamma$-bands based on the rotational aligned two-quasineutron configurations, which have
 been suggested in the framework of QRPA.

\section{ACKNOWLEDGEMENTS}
S. J. would like to acknowledge MHRD (Govt. of India) for providing the financial support to carry out the research work.   
Two of us (S. N. T. M. and J. F. S.-S.) would like to acknowledge grants from the National Research Foundation (NRF) of South Africa.
S. F. acknowledges support by the US Department of Energy grant No. DE-FG02-95ER4093.

\end{document}